\documentclass[aps,amsmath,notitlepage,twocolumn,amssymb,prl,letterpaper,longbibliography]{revtex4-1}

\usepackage{amsfonts}
\usepackage{stmaryrd}
\usepackage{bbm}
\usepackage{mathrsfs}
\usepackage{tipa}
\usepackage{amssymb}
\usepackage{txfonts}
\usepackage{graphicx}
\usepackage{dcolumn}
\usepackage{epstopdf}
\usepackage[colorlinks,linkcolor=blue,urlcolor=blue,citecolor=blue]{hyperref}
\usepackage{multirow}
\usepackage{subfigure}
\usepackage{url}

\begin{document}
\title{Revealing a Triangular Lattice Ising Antiferromagnet
in a Single-Crystal CeCd$_3$As$_3$}
\author{Y. Q. Liu$^{1}$}
\author{S. J. Zhang$^{1}$}
\author{J. L. Lv$^{2}$}
\author{S. K. Su$^{2}$}
\author{T. Dong$^{1}$}
\author{Gang Chen$^{3,4}$}
\email{gangchen.physics@gmail.com}
\author{N. L. Wang$^{1,5}$}
\email{nlwang@pku.edu.cn}
\affiliation{$^{1}$International Center for Quantum Materials, School of Physics, Peking University, Beijing 100871, China}
\affiliation{$^{2}$Institute of Physics, Chinese Academy of Sciences, Beijing 100190, China}
\affiliation{$^{3}$State Key Laboratory of Surface Physics and Department of Physics, Fudan University, Shanghai 200433, China}
\affiliation{$^{4}$Collaborative Innovation Center of Advanced Microstructures, Nanjing 210093, China}
\affiliation{$^{5}$Collaborative Innovation Center of Quantum Matter, Beijing, China}

\begin{abstract}
We report the single-crystal growth and
the fundamental magnetic and thermodynamic properties
of a rare-earth triangular lattice antiferromagnet
CeCd$_3$As$_3$. In this rare-earth antiferromagnet,
the Ce local moments form a perfect triangular lattice.
Due to the spin-orbital-entangled nature of the Ce local moments,
the compound exhibits extremely anisotropic antiferromagnetic
couplings along the c direction and in the ab plane respectively.
We show that CeCd$_3$As$_3$ represents a {\it rare} experimental
realization of an antiferromagnetic Ising model on
a two dimensional triangular lattice and thus provides
a prototype example for geometrical frustration.
We further discuss the quantum effect of the perturbative
interactions on the top of the predominant Ising interaction.
\end{abstract}

\maketitle
\begin{figure}[b]
\includegraphics[width=8cm]{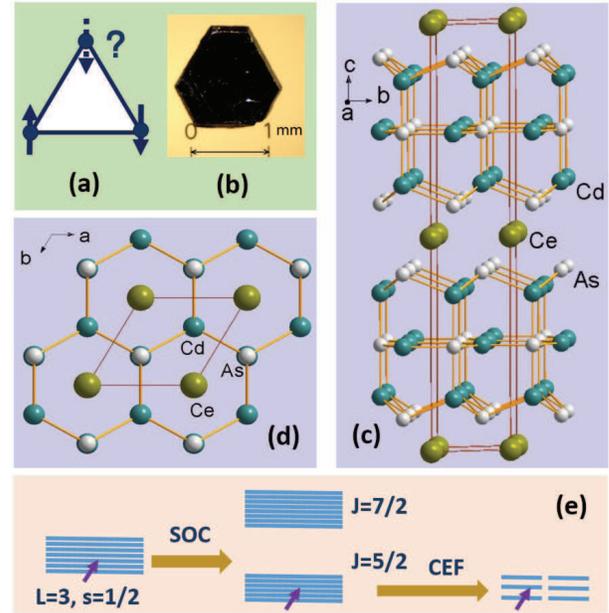}\\
\caption{
(a) A triangle of antiferromagnetically interacting
Ising spins leads to the geometrical frustration.
(b) CeCd$_3$As$_3$ single crystal grown by a vapor transport method.
(c) The crystal structure of CeCd$_3$As$_3$.
(d) The crystal structure in the ab-plane.
(e) The spin-orbit coupling (SOC) splits the Ce$^{3+}$ J=5/2 and J=7/2 states.
The 6 fold degenerate J=5/2 state is further splitted into
three Kramers doublets by the crystal electric field (CEF). }
\label{Fig:Struc}
\end{figure}

Geometrically frustrated spin systems have been a subject
of considerable theoretical and experimental interest in
modern condensed matter physics because of the potential
to host novel ground states and exotic
phenomena~\cite{Balents2010}. The prototype example of frustration
is the antiferromagnetically coupled Ising spins on
a two dimensional (2D) triangular lattice
that was first studied by Wannier~\cite{Review1950}.
When two of the spins are antiparallel arranged,
the third spin is unable to be antiparallel to both of them,
as illustrated in Fig.~\ref{Fig:Struc}(a). Then,
any spin configuration that satisfies ``two spin up one spin down"
or ``two spin down one spin up" condition in a triangle would be a
ground state. Apparently, the frustration leads to highly
degenerate classical ground state, which is expected to cause
strong quantum fluctuation and lead to various emergent quantum phenomena
when quantum spin interactions are included. Strong
quantum fluctuations, in the extreme cases, may prevent
spin ordering even at zero temperature and lead to
exotic ground states such as quantum spin liquids.

Experimental exploration of geometrically frustrated
spin systems has been made on a number of lattices
and materials, for example, on organic salts
$\kappa$-(BEDT-TTF)$_2$Cu$_2$(CN)$_3$~\cite{Yamashita2008,Yamashita2009}
and EtMe$_3$Sb[Pd(dmit)$_2$]$_2$~\cite{Yamashita2010,Itou2010}
with a ${s=1/2}$ triangular lattice, ZnCu$_3$(OH)$_6$Cl$_2$
(herbertsmithite)~\cite{Helton2007,Lee2007a}
and BaCu$_3$V$_2$O$_8$(OH)$_2$ (vesignieite)~\cite{Okamoto2009}
with a ${s=1/2}$ kagome lattice, Na$_4$Ir$_3$O$_8$ with a
${s=1/2}$ hyperkagome lattice~\cite{Okamoto2007},
and a number of rare-earth pyrochlore materials.
Quite recently, significant progress has been made in the study
of a perfect rare-earth triangular lattice antiferromagnet
YbMgGaO$_4$~\cite{Li2015c,Li2015a,Li2016g,PhysRevB.94.201114,Li2016f,Li2016c,Shen2016a,Paddison2016,Xu2016a}.
This compound was found to be disordered down to 0.048K
despite a Weiss temperature ${\Theta_{\text{W}} \approx -3}$
 K~\cite{Li2016c}. Specific heat~\cite{Li2015c,Li2015a},
$\mu$SR~\cite{Li2016c}, neutron scattering~\cite{Shen2016a,Paddison2016},
and thermal transport~\cite{Xu2016a} measurements provide
compelling evidence for the formation of a gapless U(1)
quantum spin liquid ground state in this exciting compound.
In fact, YbMgGaO$_4$ is simply the first compound that has been studied
carefully among the families of the rare-earth triangular
lattice magnets~\cite{Sanders2016a} where the strong spin-orbit
coupling of the rare-earth local moments
comes into an interplay with the geometric frustration of the underlying
lattice. In this Letter, we propose and study the magnetic properties of
another rare-earth triangular lattice antiferromagnet CeCd$_3$As$_3$
on a single-crystal sample (see Fig.~\ref{Fig:Struc} (b)).
CeCd$_3$As$_3$ belongs to a series of compounds ReA$_3$Pn$_3$
(Re = Y, La-Nd, Sm, Gd-Er; A = Cd, Zn; Pn=As, P)~\cite{Nientiedt1999,Stoyko2011,Higuchi2016,Kagayama1994},
which crystallize in the ScAl$_3$C$_3$-type structure with
the space group symmetry of P6$_3$/mmc, as displayed in
Fig.~\ref{Fig:Struc} (c) and (d). In this compound, the
lattice constants are ${a = 4.4051 \AA}$ and ${c = 21.3511 \AA}$.
The Ce$^{3+}$ ions form a perfect 2D triangular lattice,
and the Ce triangular layers are well separated from each other
by a distance of about 10.7 $\AA$ with Cd and As ions sitting in between.
Therefore it is an ideal system for studying the geometrical
frustration in a 2D triangular lattice of the Ce moments.

In CeCd$_3$As$_3$, Ce$^{3+}$ has a 4f$^1$ electron configuration
and carries a local moment. For the earth earth element,
the spin-orbit coupling is the dominant energy scale.
The atomic spin-orbit coupling entangles the orbital
angular momentum L with the spin $s=1/2$, leading to the total
angular momentum ${J=L-s=5/2}$, which has ${2J+1=6}$ fold degeneracy
on each lattice site. Under the crystal electric field with
the space group symmetry of P6$_3$/mmc, the six fold degeneracy
is further splitted into three Kramers doublets
(see Fig.~\ref{Fig:Struc} (e)). Each doublet is two fold
degenerate and is protected by time reversal symmetry.
Like the Yb$^{3+}$ ion in YbMgGaO$_4$,
the Ce$^{3+}$ mooment in CeCd$_3$As$_3$ is also described by an
effective spin ${S=1/2}$ Kramers doublet \cite{Li2016f}
if the crystal field field gap is much larger than the exchange
energy scale of the local moments.
The effective spin ${\boldsymbol S}$ for the Ce$^{3+}$ ion
involves a significant spin-orbit entanglement. As we show
below, CeCd$_3$As$_3$ represents an experimental realization
of an antiferromagnetic Ising model on a 2D triangular lattice.

The CeCd$_3$As$_3$ single crystals were grown by a vapor transport
method using Iodine as transport agent, similar to the earlier
report by Stoyko and Mar \cite{Stoyko2011}. Mixtures of Ce, Cd
and As powders with a ratio of 1:3:3 were sealed in an evacuated
quartz tube and heated at 760 $^\circ$C for a week. The mixtures
were reground and sealed together with Iodine (4 mg/liter) in a
quartz tube, which was placed into a two stage horizontal furnace.
The hot end with the mixture was heated at 800 $^\circ$C and the
cold end at 700 $^\circ$C for two week. Single crystals of
CeCd$_3$As$_3$ (ScAl$_3$C$_3$-type) were obtained in the form
of hexagonal plate-shape (see the picture in Fig.~\ref{Fig:Struc} (b)).
Energy-dispersive X-ray (EDX) analysis with a scanning electron
microscope (SEM) on the crystals revealed chemical compositions
of Ce:Cd:As=1:3:3.

\begin{figure}[htbp]
  \centering
  \includegraphics[width=6cm]{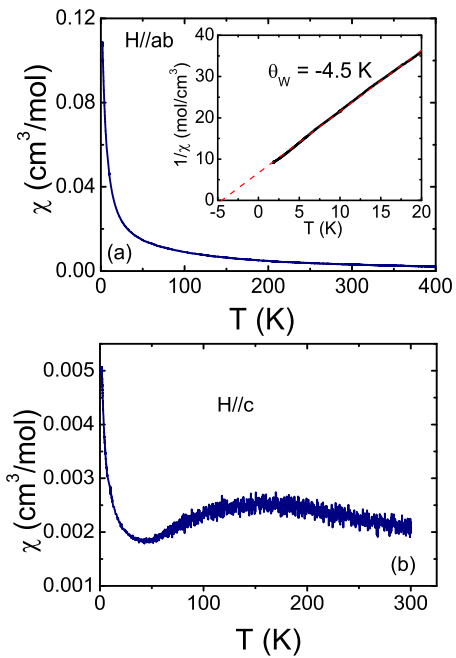}\\
  \caption{The temperature dependent magnetic susceptibilities
  measured at a field of 1T with \textbf{H} $\parallel$ ab plane
  (a) and \textbf{H} $\parallel$ c-axis (b) for the CeCd$_3$As$_3$
  single crystal. Inset of (a): the inverse susceptibility below 20 K
  together with fit by the Curie-Weiss law (in red dash line).}\label{Fig:MT}
\end{figure}

Magnetization measurements were performed using a
Quantum Design physical property measurement system
(PPMS) in the temperature range of 1.8 - 400 K under
0 - 16 T. Figure \ref{Fig:MT} shows the temperature
dependent magnetic susceptibilities measured at a
field of 1T with \textbf{H} parallel and perpendicular
to the triangular plane, respectively. For H $\parallel$ ab-plane,
the magnetic susceptibility displays a Curie-Weiss like behavior.
There is no indication of magnetic order down to the
lowest measured temperature of 1.8 K. As we shall see
below from the specific heat measurement in the He3
temperature range, no sign of magnetic order was seen
down to 0.6 K. The inverse of the susceptibility below
20 K shown in the inset of Fig. \ref{Fig:MT} (a) could
be well fitted by the Curie-Weiss law with an antiferromagnetic Weiss
temperature $\theta_{\text W}=-4.4$ K.
According to the generic Hamiltonian
proposed for the rare-earth triangular systems in earlier
studies on YbMgGaO$_4$ \cite{Li2015c,Li2015a} (also see Eq.~\ref{eq1} below),
the Weiss temperature is related to the antiferromagnetic
interaction by ${\theta_{\text W}=-3 J_{\pm}}$~\cite{Li2015a,Li2016f},
then we obtain in-plane antiferromagnetic coupling
$J_{\pm}\approx1.5 $ K. As a comparison,
$J_{\pm}\approx0.9 $ K for YbMgGaO$_4$~\cite{Li2015a}.

On the other hand, when the field is applied parallel to the c axis,
the susceptibility shows significantly different behavior.
As shown in Fig. \ref{Fig:MT} (b), a broad peak is observed
near 150 K. Such broad peak is commonly seen in 1D or 2D systems
with strong antiferromagnetic interactions~\cite{Shimizu2003}.
The upturn below 40 K in the susceptibility could be ascribed
to the magnetic defects or impurities in the sample.
The result provides clear evidence for the presence of strong
antiferromagnetic coupling along the c-axis. In general,
one expects that antiferromagnetic correlation of any sort
would depress susceptibility, because the correlations
resist the alignment of spins with the applied field.
As temperature increases, the antiferromagnetic correlations
are expected to decrease, leading to an increase of magnetic
susceptibility. At high temperature when the antiferromagnetic
correlation is weak enough and the moments are effectively
decoupled, then a Curie-Weiss behavior is expected to occur.
As a result, a peak would develop in the magnetic susceptibility.
The location of the peak roughly reflects the energy scale of
the antiferromagnetic coupling $J_{zz}$ along c-axis.
For instance, for an s=1/2 2D square lattice antiferromagnet,
$J=k_BT_{max}/0.9$ \cite{Shimizu2003,PhysRevLett.61.2971}.
Another possibility for the broad peak in the susceptibility
is the thermal excitations of the excited
Kramers' doublets. This effect has been observed, for instance,
in Sm$_2$Ti$_2$O$_7$~\cite{SmTiOa,SmTiOb}. Whether similar
effect is relevant for CeCd$_3$As$_3$ is not quite obvious, it is
likely both the activation of excited doublets and
strong $J_{zz}$ interaction may be present. The following
magnetization measurement with increasing fields further supports
a large anisotropy for the antiferromagnetic spin interaction
in the ab plane and along the c axis.


\begin{figure}[htbp]
\centering
\includegraphics[width=6.5cm]{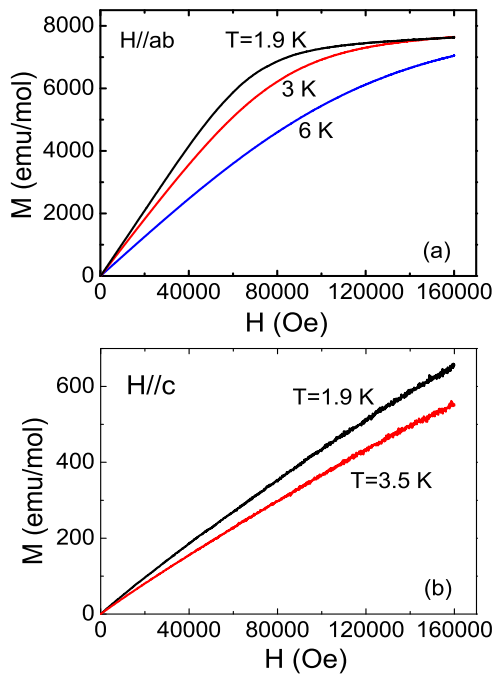}\\
\caption{The field dependent magnetization measurements
up to 16 T at several selected temperatures for
\textbf{H} $\parallel$ ab plane (a) and \textbf{H} $\parallel$ c-axis
(b) for CeCd$_3$As$_3$ single crystal, respectively.}
\label{Fig:MH}
\end{figure}

In Fig.~\ref{Fig:MH}, we show the field dependent magnetizations up
to 16 T at several selected temperatures with \textbf{H} parallel
and perpendicular to the triangular plane, respectively.
For H $\parallel$ ab plane (Fig.~\ref{Fig:MH} (a)), the
magnetization increases linearly with field below 5 T,
then tends to saturate at higher magnetic field. When
the field is higher than 12 T, the magnetization at
low temperature (e.g. at 1.9 K) becomes linearly dependent
on the field with a very small slope, which is understood
as the Van Vleck susceptibility. The result is similar
to that seen in YbMgGaO$_4$~\cite{Li2015c,Li2015a}, except
for the slightly higher magnetic field for the magnetization
saturation. The observation also suggests a bit higher in-plane
antiferromagnetic coupling strength for CeCd$_3$As$_3$ than
YbMgGaO$_4$. By contrast, for H$\parallel$c-axis (Fig.~\ref{Fig:MH} (b)),
the magnetization shows roughly a linear increase with
field up to the highest measurement field (16 T) without
showing any tendency towards a saturation. This is because
the antiferromagnetic interaction is rather strong for
c-axis direction, a magnetic field of 16 T is still
too small as compared to the c-direction coupling strength
$J_{zz}$.
We thus conclude from the susceptibility and magnetization measurements
that the antiferromagnetic coupling along the c-axis $J_{zz}$
may be one or two orders larger than the in-plane $J_{\pm}$.

\begin{figure}[htbp]
  \centering
  \includegraphics[width=6.5cm]{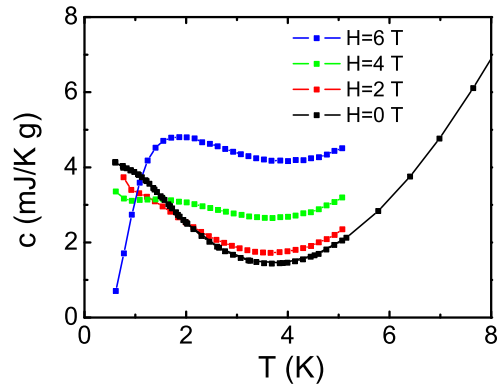}\\
  \caption{The low temperature specific heat data measured
  down to 0.6 K. No magnetic order is indicated. With applying
  magnetic field the specific heat values increases above 1.5 K,
  and a broad hump is clearly observed for H $=6$ T. The magnetic
  field induced change suggests an absence of magnetic order
  even below our lowest measurement temperature. }\label{Fig:SH}
\end{figure}

To demonstrate whether the spins of CeCd$_3$As$_3$
become ordered at further lower temperature, we performed
specific heat measurement with a He3 cryostat in PPMS system.
Fig.~\ref{Fig:SH} shows the specific heat data down to 0.6 K.
The specific heat at zero field decreases with decreasing
temperature and reaches a minimum near 3.5 K, then shows
an upturn at lower temperature. The data are very similar
to that seen for YbMgGaO$_4$~\cite{Li2015c}, except that
the minimum appears at lower temperature for CeCd$_3$As$_3$.
In YbMgGaO$_4$ the minimum locates near 10 K at zero field,
below which the specific heat increases and forms a broad
hump at 2.4 K. The broad hump is an indication for the
crossover into a quantum spin liquid state~\cite{Yamashita2008,Li2015c}.
Below the hump the magnetic specific heat would
follow a power-law temperature dependence. In our current
measurement, we could not access to the temperature
below 0.6 K, so the downward turn below 0.6 K is not
yet observed for CeCd$_3$As$_3$. Nevertheless, the
compound does not show any magnetic order down to our
lowest measured temperature.

We also performed the specific heat measurement under the
external magnetic field parallel to the ab-plane. The data are also shown in
Fig.~\ref{Fig:SH}. If there exists an antiferromagnetic
order, i.e. presence of a sharp $\lambda$-shape peak, below
our lowest measured temperature of 0.6 K, we expect that
the magnetic field would suppress the values of specific heat
above the ordering temperature and shifts the $\lambda$-shape
peak to lower temperature. However, applying the magnetic field
actually increases the value of specific heat at high temperature
above 1.5 K. A broad hump is clearly visible for
a magnetic field of 6 T. The result is again very similar to
the case of YbMgGaO$_4$ for which the applied magnetic field
induces a shift of the broad hump to a higher temperature.
The field dependent measurement may indicate an
absence of magnetic order even below our lowest
measured temperature.

Now let us discuss the implication of the experimental data.
Due to the spin-orbit-entangled nature of the Kramers doublets,
the interaction between the effective spin-1/2 moments is highly
anisotropic. Based on earlier works, the most generic spin
Hamiltonian allowed by the space group symmetry of the
rare-earth triangular system is given
by~\cite{Li2015a,Li2016g,PhysRevB.94.201114,Li2016f}
\begin{eqnarray}
{\mathcal H} &=&
\sum_{\langle ij \rangle}
  J_{zz}   S_i^z S_j^z
+ J_{\pm} (S^+_i S^-_j + S^-_i S^+_j)
\nonumber \\
&+&  J_{\pm\pm}
(\gamma_{ij} S^+_i S^+_j
+ \gamma_{ij}^{\ast} S^-_i S^-_j)
\nonumber \\
&-&  \frac{i J_{z\pm}}{2}\big[
 ({\gamma_{ij}^{\ast} S^+_i
- \gamma_{ij} S^-_i}) S^z_j
+S^z_i
 (\gamma_{ij}^{\ast} S^+_j
- \gamma_{ij} S^-_j) \big],
\label{eq1}
\end{eqnarray}
where $S_i^\pm = S_i^x \pm i S_i^y$, and $\gamma_{ij} =
\gamma_{ji} = 1, e^{i2\pi/3}, e^{-i2\pi/3}$ are the phase factors
for the bond $ij$ along three 120$^\circ$ directions, respectively.
The first line of Eq.~(\ref{eq1}) is the standard XXZ model.
The $J_{\pm}$ and $J_{zz}$ are the antiferromagnetic interactions
within and perpendicular to the triangular layer.
The $J_{\pm\pm}$ and $J_{z\pm}$ terms of Eq.~(\ref{eq1})
define the anisotropic interactions that arise from the strong
spin-orbit coupling. According to the experiments presented above,
both $J_{\pm}$ and $J_{zz}$ $ > 0$, i.e. antiferromagnetic coupling,
more significantly, $J_{zz} \gg J_{\pm}$. Therefore,
the system is in the Ising limit with $J_{\pm}$ as a perturbation.
In the degenerate Ising ground state manifold, the perturbative
effect of the $J_{\pm\pm}$ and $J_{z\pm}$ terms only appears at
the second order and can be ignored if they are not so signicant
compared to $J_{zz}$. Therefore, we expect the antiferromagnetic
Ising model to be the starting point to understand the properties
of CeCd$_3$As$_3$. Numerous theoretical
studies of antiferromagnetic Ising model and
the related ones on the triangular lattice
over the last several decades, however, we are not aware
of any real material most closely corresponding to the model.
CeCd$_3$As$_3$ may represent the first material realization
of antiferromagnetic Ising model on the triangular lattice.

\begin{figure}[htbp]
\centering
\includegraphics[width=7cm]{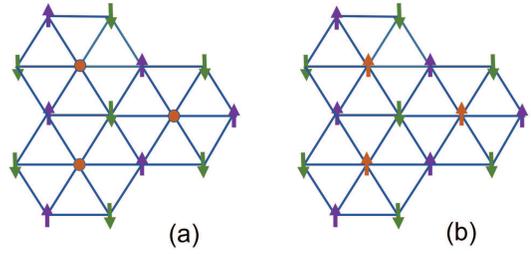}\\
\caption{(a) The three-sublattice ordered structures in the ground state
for the antiferromagnetic Ising model on the triangular lattice with
presence of a transverse field~\cite{Sondhi}.
(b) A spin configuration with two of the
sublattices showing up-spins and the remaining down.
It corresponds to a 1/3-plateau of the saturation value in
magnetization. }
 \label{Fig:Magpattern}
\end{figure}

The frustration in the triangular lattice may cause various
exotic properties. As mentioned in the introduction, the ground
state of antiferromagnetic Ising model on the triangular lattice
is highly degenerate and has critical spin correlation.
In the thermodynamic limit, the system has
a finite entropy even at T = 0 K~\cite{Review1950}. With presence
of $J_{\pm}$ as a perturbation, both the c-direction component and
ab-plane component would develop three-sublattice spin structures
in the ground state~\cite{PhysRevLett.112.127203,PhysRevB.91.081104}.
Since $J_{\pm}$ is rather small, the three-sublattice magnetic order
may possibly occur at a a temperature that is much lower than the
measured ones in this work. If one applies a magnetic field in the
ab plane, one obtains a transverse field quantum Ising model on a
triangular lattice. It is known that the system develops a 3-sulattice
order via a quantum order by disorder mechanism in the low field
regime~\cite{Sondhi}. The two of them are antiferromagnetic coupled
in a hexagonal ring, while the third one in the center of the ringed
spins is alligned with the field (see Fig.~\ref{Fig:Magpattern} (a) for an illustration).
If the field is applied along the c-axis to the Ising
spin system, one would expect to observe a 1/3 plateau of
the saturation value in the magnetization,
which corresponds to the spin configuration of two of the
sublattices have up-spins and the remaining down (see
Fig.~\ref{Fig:Magpattern} (b)) \cite{doi:10.1143/JPSJ.55.4448}.
Future measurement under high magnetic field is certainly
desired in order to verify the scenario.

Finally, we would like to point out that
the rare-earth triangular, kagome, fcc, and pyrochlore materials~\cite{Sanders2016a,Nientiedt1999,Stoyko2011,li2016kitaev,PhysRevB.86.104412,li2016octupolar,Onoda2010,Ross2011,TbTiO,PhysRevLett.108.037202,0034-4885-77-5-056501,Li2015c,Li2015a,Huang2014,Li2016g,PhysRevB.94.201114,Li2016f,Li2016c,Shen2016a,Paddison2016,Xu2016a,Curnoe2008,
Higuchi2016,Kagayama1994,PhysRevB.94.205107,PhysRevB.88.144402,Sanders2016,Dun2016,PrZrO2013},
where the spin-orbit entanglement of the local moments
meets with strong geometrical frustration,
represent an interesting new direction to explore quantum magnetism
with anisotropic spin interactions and exotic phenomena.
Even for the very similar trianglular lattices,
YbMgGaO$_4$ has relatively close antiferromagnetic
coupling strengths along c-axis and ab-plane,
leading to a gapless U(1) quantum spin liquid ground state,
whereas CeCd$_3$As$_3$ shows extremely anisotropic
antiferromagnetic coupling strengths, realizing a 2D
Ising antiferromagnet on a triangular lattice.
Another difference is that the antiferromagnetic coupling
$J$ is very small for YbMgGaO$_4$ (both $J_{zz}$ and $J_{\pm}$ $\sim$1 K),
while for CeCd$_3$As$_3$, the antiferromagnetic coupling along
c-direction $J_{zz}$ is considerably larger.
Since the spins are not ordered down to at least 0.6 K,
the compound represents an extremely frustrated system.
In fact, rare earth geometrical frustration structures
can be realized in many different compounds, for example,
in the CaAl$_2$Si$_2$-type trigonal structure with
a $P\overline{3}m1$ space group~\cite{PhysRevB.94.045112},
besides a few structures mentioned in the introduction.
Exploration of quantum magnetism in those different
systems would be of much interest.

\section{Acknowledgement}

We acknowledge experimental help from Peijie Sun and useful
discussions with Xincheng Xie. This work was
supported by the National Science Foundation of China (No.~11327806),
and the National Key Research and Development Program of
China (No.~2016YFA0301001, No.~2016YFA0300902).

\bibliographystyle{apsrev4-1}
\bibliography{CeCd3As3}

\end{document}